\documentclass[runningheads]{llncs}
\usepackage[T1]{fontenc}
\usepackage{graphicx,verbatim}
\usepackage{subcaption}
\usepackage{amsmath, amssymb, algorithm, algpseudocode}
\usepackage{amsmath}
\usepackage{multirow}
\usepackage{amssymb}
\usepackage{bm}
\usepackage{ulem}
\usepackage[colorlinks]{hyperref}
\usepackage[table,xcdraw]{xcolor} 

\usepackage{cleveref}
\usepackage{booktabs}
\usepackage{marvosym}
\begin{document}
\title{U-RWKV: Lightweight medical image segmentation with direction-adaptive RWKV}
\titlerunning{U-RWKV}
%

\authorrunning{H. Ye et al.}

\author{Hongbo Ye\inst{1,2} 
\and Fenghe Tang\inst{1,2} 
\and Peiang Zhao \inst{1,2} 
\and Zhen Huang\inst{1,2} 
\and Dexin Zhao\inst{1,2} 
\and Minghao Bian\inst{1,2} 
\and S. Kevin Zhou\inst{1,2,3,4}$^{\href{mailto:skevinzhou@ustc.edu.cn}{\textrm{\Letter}}}$ 
}

\institute{School of Biomedical Engineering, Division of Life Sciences and Medicine, University of Science and Technology of China (USTC), Hefei Anhui, 230026, China \and
Center for Medical Imaging, Robotics, Analytic Computing \& Learning (MIRACLE), Suzhou Institute for Advance Research, USTC \and
Jiangsu Provincial Key Laboratory of Multimodal Digital Twin Technology, Suzhou
\and
State Key Laboratory of Precision and Intelligent Chemistry, USTC}
    
\maketitle              
%

\begin{abstract}

Achieving equity in healthcare accessibility requires lightweight yet high-performance solutions for medical image segmentation, particularly in resource-limited settings. Existing methods like U-Net and its variants often suffer from limited global Effective Receptive Fields (ERFs), hindering their ability to capture long-range dependencies. To address this, we propose U-RWKV, a novel framework leveraging the Recurrent Weighted Key-Value(RWKV) architecture, which achieves efficient long-range modeling at O(N) computational cost.
The framework introduces two key innovations: the Direction-Adaptive RWKV Module(DARM) and the Stage-Adaptive Squeeze-and-Excitation Module(SASE). DARM employs Dual-RWKV and QuadScan mechanisms to aggregate contextual cues across images, mitigating directional bias while preserving global context and maintaining high computational efficiency. SASE dynamically adapts its architecture to different feature extraction stages, balancing high-resolution detail preservation and semantic relationship capture.
Experiments demonstrate that U-RWKV achieves state-of-the-art segmentation performance with high computational efficiency, offering a practical solution for democratizing advanced medical imaging technologies in resource-constrained environments. The
code is available at \url{https://github.com/hbyecoding/U-RWKV}.


 

\end{abstract}

\section{Introduction}
\label{sec:intro}

\begin{figure}[t]
    \centering
    \begin{subfigure}{0.47\textwidth}
        \centering
        \includegraphics[width=\linewidth]{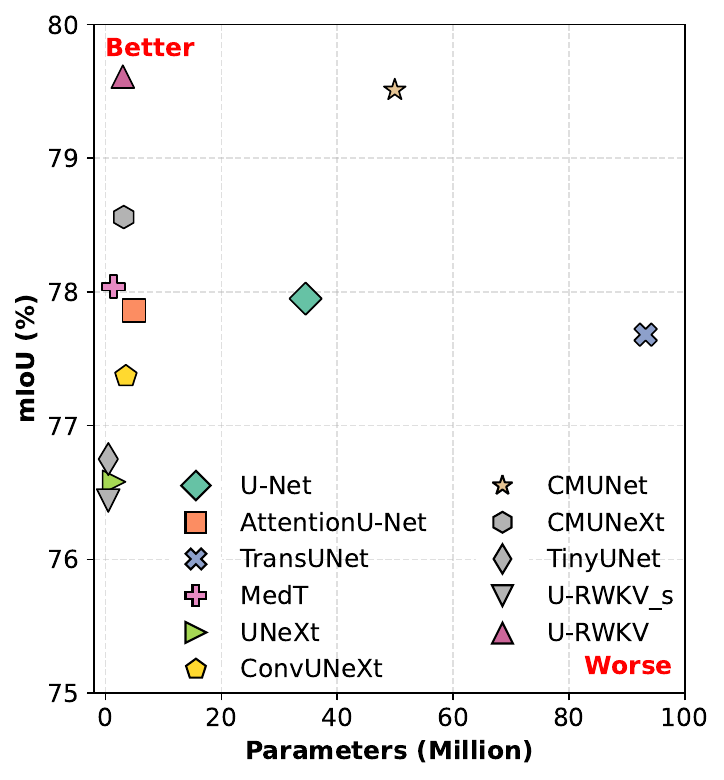}
        \caption{\small U-RWKV achieves the highest Avg. Dice with efficient parameters}
        \label{fig:performance}
    \end{subfigure}
    \hfill
    \begin{subfigure}{0.46\textwidth}
        \centering
        \includegraphics[width=\linewidth]{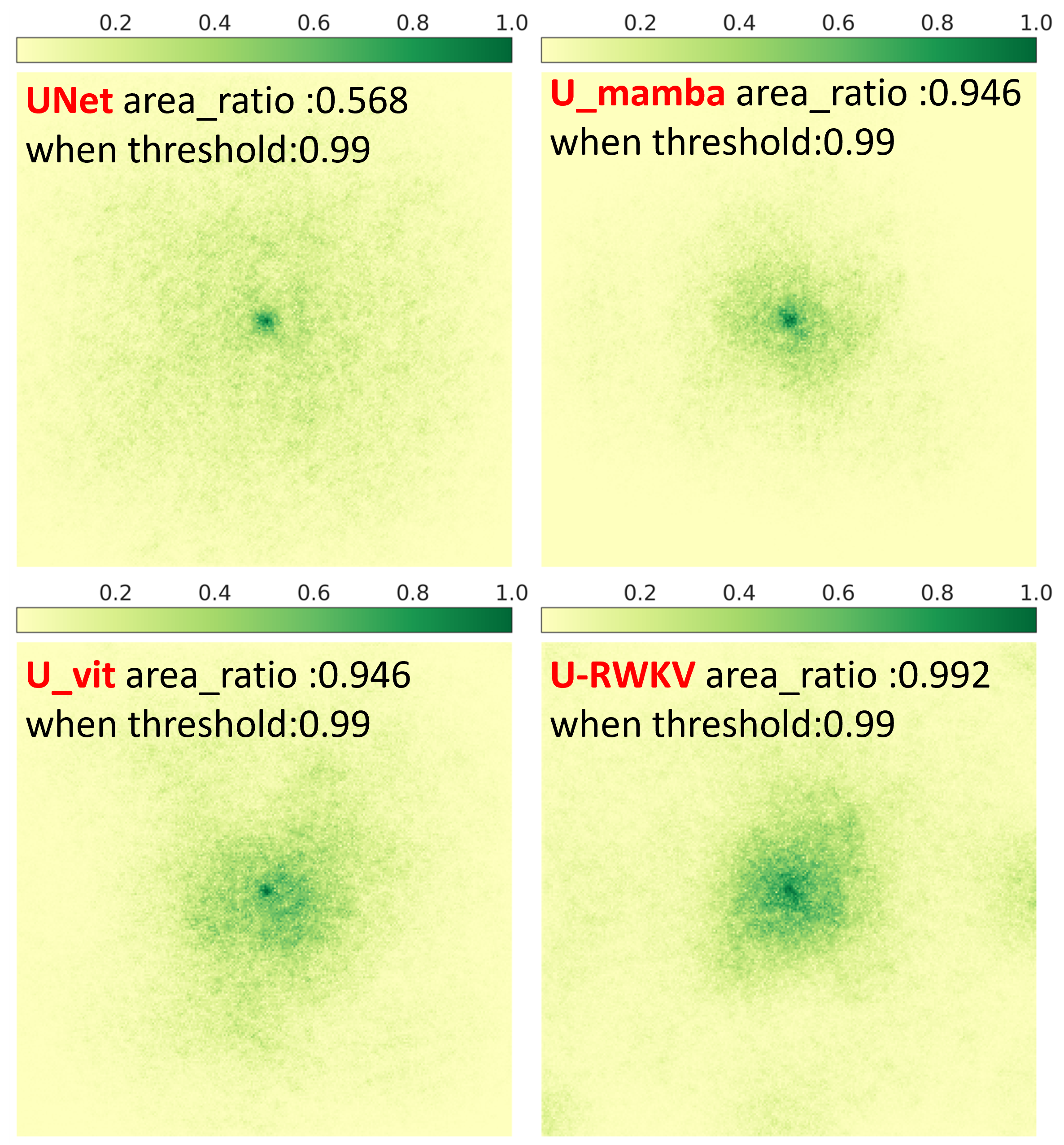}
        \caption{\small U-RWKV has a larger ERF than U-Mamba and U-ViT-please see~\ref{AblationStudies}}
        \label{fig:erf}
    \end{subfigure}
    \caption{Performance comparison and Effective Receptive Field (ERF).}
    \label{fig:ERF_and_scatter}
\end{figure}

Bridging the gap in healthcare accessibility requires not only breakthroughs in medical technology but also solutions that can be widely deployed across diverse clinical environments, especially in resource-limited settings~\cite{richardson2022nature}.
In the field of medical image segmentation, while convolutional neural networks (CNNs), such as U-Net~\cite{ronneberger2015unet} and its variants~\cite{oktay2018attentionunet,zhou2018unet++,huang2020unet3plus,tang2023cmu,han2022convunext,valanarasu2022unext,tang2024cmunext,tang2023mobileutr}, have achieved initial success through localized feature extraction, they fundamentally suffer from inadequate global Effective Receptive Fields(ERFs)~\cite{luo2016understanding,ding2022scaling}, as shown by U-Net's 0.568 high-contribution ratio at 0.99 threshold (Fig.~\ref{fig:ERF_and_scatter}(b)). Lightweight yet high-performing models with global ERF thus hold immense potential, offering a viable and equitable pathway to democratize access to advanced medical imaging technologies.

To address the limitations of existing methods~\cite{chen2021transunet,cao2022swin,valanarasu2021medical,Jhadoubleunet2020,jha2019resunet++,hz2-huang2024pele,hz3-zhou2024hybrid}, we propose \textbf{U-RWKV}, a novel lightweight framework that leverages the emerging Recurrent Weighted Key-Value (RWKV) architecture~\cite{peng2023rwkvcem}. RWKV achieves long-range modeling at $O(N)$ computational cost, offering a powerful foundation for efficient medical image segmentation through its linear-complexity attention mechanism.
At the core of our U-RWKV are two key components: the Direction Adaptive RWKV Module (DARM) and the Stage Adaptive Squeeze and Excitation Module (SASE). These modules work synergistically to model long-range spatial dependencies while maintaining computational efficiency,
setting it apart from traditional transformer-based or convolutional models, as show in Fig.~\ref{fig:ERF_and_scatter}(a).

The DARM is designed to dynamically aggregate contextual cues across the entire image by introducing two innovative mechanisms: \textbf{Dual-RWKV} and \textbf{QuadScan}. The core algorithm of RWKV, inspired by RNN-like WKV computations, is inherently designed for processing one-dimensional sequential data. However, this presents a challenge when adapting it to visual data, which lacks an inherent sequential arrangement of components. To address this issue, we propose the {Dual-RWKV} mechanism, which processes 2D feature maps as dual 1D sequences—one in the original order and the other in reverse order. This bidirectional design ensures cross-orientation context preservation while eliminating directional bias. By propagating information bidirectionally, Dual-RWKV captures multi-scale contextual dependencies, mitigating information loss in ambiguous regions such as diffuse boundaries or corner-situated lesions.

On the other hand, the complex spatial relationships and diverse modalities (e.g., CT, MRI) in medical imaging demand adaptive mechanisms capable of capturing anisotropic features~\cite{tang2024hyspark,tang2025mambamim,tang2025hi}. To meet this requirement, we introduce \textbf{QuadScan}, a quad-directional scanning strategy that traverses the image through four directional flows: left→right, right→left, top→bottom, and bottom→top. Each image patch acquires contextual knowledge exclusively through a compressed hidden state computed along its corresponding scanning path, reducing computational complexity while preserving global context. This systematic integration of edge semantics from multiple directions achieve global ERF, as shown in Fig.~\ref{fig:ERF_and_scatter}(b). 

To further enhance the adaptability of U-RWKV, we introduce the Stage-Adaptive Squeeze-and-Excitation Module (SASE). SASE dynamically adjusts its architecture based on the stage of feature extraction. In early stages, SASE employs dilated inverted bottleneck structures to preserve high-resolution features, ensuring detailed spatial information is retained. In deeper layers, SASE transitions to compact bottleneck designs to maintain computational efficiency while capturing high-level semantic relationships. This stage-adaptive design enables U-RWKV to generalize effectively across different datasets, accommodating the intricate spatial correlations and semantic relationships inherent in medical imaging modalities such as CT and MRI.

In summary, our main contributions are as follows: \textbf{(I)} We propose \textbf{U-RWKV}, a lightweight framework balancing computational efficiency and segmentation performance for resource-constrained settings; \textbf{(II)} We introduce two innovations: (a) \textbf{DARM}, which uses \textbf{Dual-RWKV} and \textbf{QuadScan} to model long-range dependencies efficiently while reducing directional bias; and (b) \textbf{SASE}, which adapts dynamically to enhance the model's robustness and generalization; \textbf{(III)} Comprehensive experiments validate U-RWKV's state-of-the-art performance, efficiency, and adaptability across diverse medical imaging tasks.

\section{Method}
\label{sec:method}

\subsection{Architecture Overview}
\label{subsec:architecture}

\begin{figure}[!t]
    \centering
    \includegraphics[width=0.8 \linewidth]{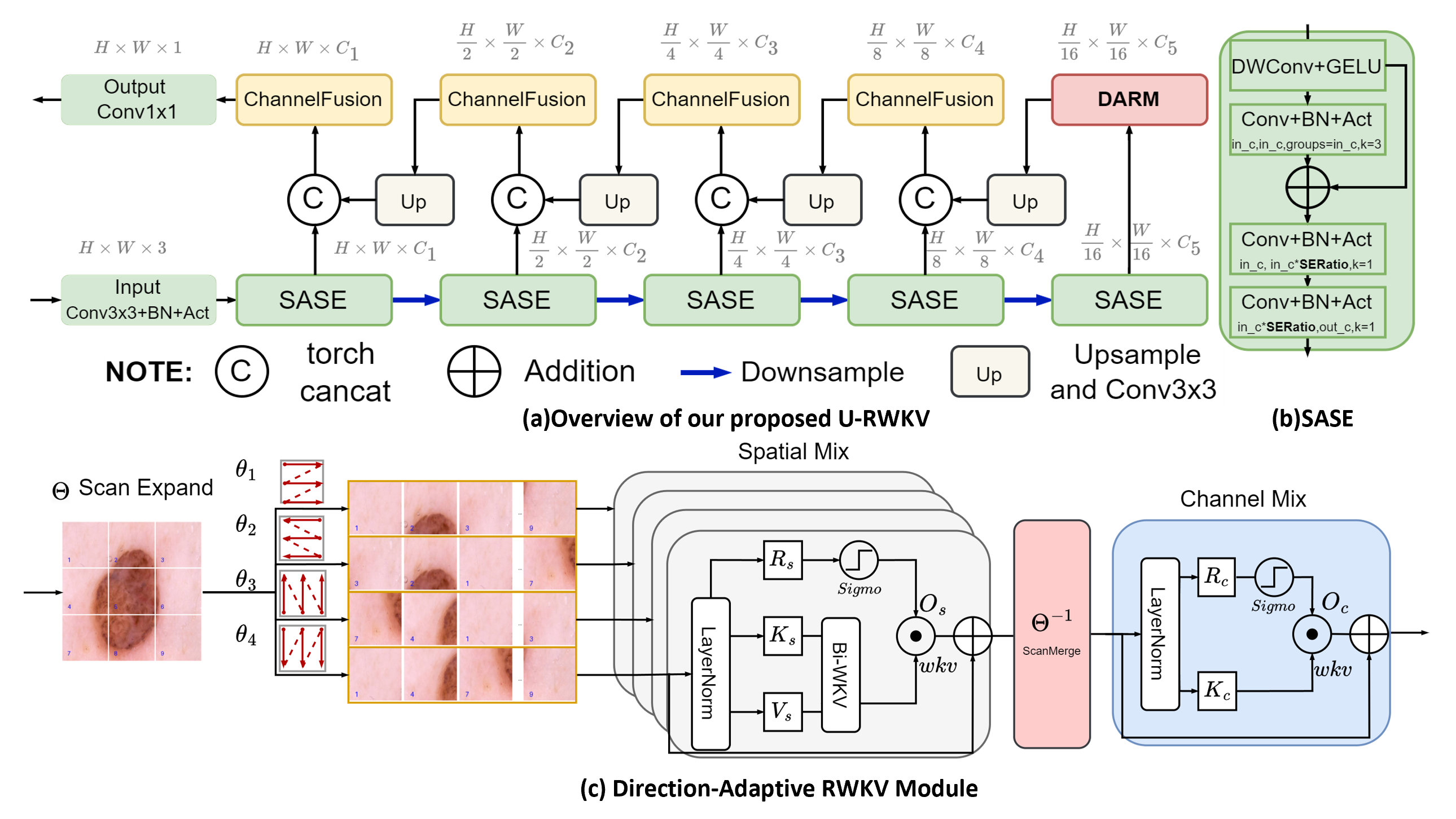}
    \caption{Overview of the proposed U-RWKV; SASE Module and DARM}
    \label{fig:architecture}
\end{figure}

The proposed architecture introduces a novel U-shaped encoder-decoder framework tailored for medical image segmentation, as depicted in Fig.~\ref{fig:architecture}(a). The model is designed to efficiently process input images through a hierarchical structure that captures multi-scale features. The encoder progressively reduces the spatial dimensions of the input while increasing the channels. It uses a series of convolutional layers with $3 \times 3$ kernels and stride=2 for downsampling. The decoder path aims to reconstruct the feature maps through a series of upsampling operations, performed by transposed convolutions that gradually restore the spatial resolution. This process is named \textit{ChannelFusion} because it involves two layers of CNNs. Each convolutional layer is followed by batch normalization to ensure robust feature extraction and regularization.

\subsection{Stage-Adaptive SE Module (SASE)}
The \textit{SASE} block dynamically adapts to different stages of the network to complement RWKV. Its architecture enhances the hierarchical feature transmission to DARM. Specifically, this design uses lightweight pointwise convolutions and inverted residual structures in shallow modes. Shallow mode refers to early stages where the resolution is high and the feature channel ratio, calculated as the number of channels divided by the product of height and width, is large. In these stages, the SERatio is set to 4, meaning the output channels are four times the input channels. In deeper stages, where the ratio of channels to H×W is smaller (C/H/W $\geq$  1), we perform channel-wise splitting into 8 parts and then double the channels, making the SERatio effectively 1/4. We also use depthwise separable convolutions in these deeper modes to improve spatial feature extraction. Using lightweight convolutions and residuals in shallow stages balances efficiency and feature richness. This resolution-aware design enhances the feature informativeness of DARM.

\subsection{Direction-Adaptive RWKV Module (DARM)}  
\label{subsec:darm}  

As discussed earlier, the fine-grained local features extracted by the encoder need to incorporate long-range dependencies to enable effective fusion of local and global information in the decoder. To achieve this, we propose DARM, which refines the encoded features while preserving their spatial and channel-wise relationships, leveraging the temporal and channel mixers from RWKV and Vision RWKV.  

\subsubsection{Preliminaries: RWKV for vision data}  

RWKV processes an input $s \in \mathbb{R}^{T \times C}$, where $C$ is the number of channels and $T$ is the sequence length. First, layer normalization (LN) stabilizes the features. The normalized features are projected into three components: receptance $R_s$, key $K_s$, and value $V_s$, via learnable matrices $W_R, W_K, W_V \in \mathbb{R}^{C \times C}$.  

Borrowing from ~\cite{duan2024vision}, we define spatial and channel mixing as follows:  
\textbf{Spatial mixing (\text{spa(·)})} is a token-wise aggregation:  
\begin{equation}  
spa_t = wkv_t = \operatorname{Bi-WKV}(K_s, V_s)_t = \frac{\sum_{i=1, i \neq t}^{T} e^{-\frac{|t-i|-1}{T} \cdot w + k_i} v_i + e^{u + k_t} v_t}{\sum_{i=1, i \neq t}^{T} e^{-\frac{|t-i|-1}{T} \cdot w + k_i} + e^{u + k_t}},  
\label{eq:bi-wkv}  
\end{equation}  

where $u$ and $w$ are learnable parameters controlling local and non-local interactions respectively. This enables dynamic importance adjustment of nearby and distant tokens, ensuring robust feature aggregation.  

\textbf{Channel mixing (\text{cha(·)})} is a pointwise feed-forward network applied across the channel dimension.  

For vision data $X \in \mathbb{R}^{C \times H \times W}$, we transform spatial features into a sequence using \text{Vision2Seq(·)}, which is a more sophisticated process than simply reading each row sequentially (\text{Flatten(·)}). The VRWKV process then combines spatial and channel mixing:  
\begin{equation}  
\text{VRWKV}(X) = \text{cha} (\text{Flatten}( \text{spa} ( \text{Vision2Seq}(X) ) ) ).  
\label{eq:RWKV}  
\end{equation}  

\noindent For the DARM input, let $E \in \mathbb{R}^{C \times H \times W}$ be the encoded features and $D_{\text{pre}} \in \mathbb{R}^{C \times H \times W}$ the refined features. Since $H$ and $W$ are reduced due to encoder downsampling, we set the patch size in DARM's patch embedding to 1, and the sequence length to $H \times W$. As shown in Fig.~\ref{fig:architecture}(c), $E$ first passes through the \textbf{QuadScan} mechanism, which operates along spatial dimensions while keeping channel information intact.  

\begin{algorithm}[t!]  
\caption{DARM: Direction-Adaptive RWKV Module}  
\label{alg:darm}  
\begin{algorithmic}[1]  
\Require $E \in \mathbb{R}^{C \times H \times W}$ (encoder features)  
\Ensure $D_{\text{pre}} \in \mathbb{R}^{C \times H \times W}$ (features for decoder)  

\State \textbf{Step 1: QuadScan Expansion and Spatial Mix}  
\For{each direction $ i \in \{L \rightarrow R,\ R \rightarrow L,\ T \rightarrow B,\ B \rightarrow T\} $}  
    \State $ s_i = \theta_i(E) $ \Comment{Apply directional scan $\theta_i: \mathbb{R}^{C \times H \times W} \to \mathbb{R}^{T \times C}$}  
    \State $ s_i' = \text{spa}(s_i) $ \Comment{Spatial mix}  
\EndFor  

\State \textbf{Step 2: QuadScan Merge}  
\State $ E_i = \theta_i^{-1}(s_i') $ \Comment{Reconstruct 2D features from sequences}  
\State $ E' = \text{Average}(\text{Stack}(E_1, E_2, E_3, E_4)) $ \Comment{Pixel-wise averaging}  

\State \textbf{Step 3: Channel Mixing}  
\State $ D_{\text{pre}} = \text{cha}(\text{Flatten}(E')) $ \Comment{Channel-wise enhancement}  

\State \textbf{Return} $ D_{\text{pre}} $  
\end{algorithmic}  
\end{algorithm}  

\noindent \textbf{QuadScan Mechanism:} This operation scans the feature map \(E\) along four directions: left-to-right, right-to-left, top-to-bottom, and bottom-to-top. Each directional scan produces a 1D sequence \(s_i\) via \(\theta_i\). These sequences undergo spatial mixing through \text{spa(·)} to refine long-range dependencies. After processing, the inverse functions \(\theta_i^{-1}\) reconstruct spatial features \(E_i\), which are then averaged pixel-wise into the final feature map \(E'\). The resulting feature map is subsequently flattened into a sequence and sequentially fed into the channel mix module, enabling the model to capture a comprehensive receptive field for subsequent processing.  

\noindent\textbf{Dual-RWKV Mechanism.} This core feature refinement module processes 2D feature maps as two separate 1D sequences—one in the original order and the other in reverse—without weight sharing between directions. This bidirectional design preserves cross-orientation context while preventing directional bias. By propagating information in both forward and backward passes independently, Dual-RWKV captures richer spatial dependencies. When combined with QuadScan, the model can capture complementary directional information, further enhancing its robustness and adaptability. 

Let the symbols be defined as above. The unified process of our \textbf{Direction-Adaptive RWKV Module (DARM)} can be formulated as:
\begin{equation}
\small
{D_{pre}}=\text{cha}\left(\text{Flatten}\left(\Theta^{-1}\left(\text{spa}\left( s\right)\right)\right)\right)+\text{cha}\left(\text{Flatten}\left(\Theta^{-1}\left(\text{spa}\left(s^{\leftarrow})\right)\right)\right)\right)
\end{equation}

\section{Experiments and Results}
\label{sec:experiments}

\subsection{Settings}
\label{subsec:datasets}
\textbf{Datasets.} Our study utilizes diverse datasets. The BUSI dataset~\cite{al2020dataset} consists of breast ultrasound images from 600 female patients, with 780 images in total, classified into normal, benign, and malignant. Kvasir~\cite{guo2020polyp} and ClinicDB~\cite{bernal2015wm} are polyp-related endoscopic datasets. Kvasir has 1,000 manually-annotated polyp images, and ClinicDB contains 612 static images from colonoscopy videos. The ISIC 2017 and 2018 datasets~\cite{codella2019skin} focus on skin diseases, with different numbers of training and test images.

\noindent\textbf{Metrics.} In medical image segmentation, we commonly use the Dice Similarity Coefficient (DSC) and the Intersection over Union (IoU) to evaluate performance. Higher values of DSC and IoU indicate better segmentation accuracy.

\noindent\textbf{Implementation Details.} 
The training procedure follows the settings described in ~\cite{tang2024cmunext,tang2023cmu}, with the following modifications: training is conducted for 280 epochs on a single NVIDIA 3090 GPU; the official Synapse dataset is used exclusively, while for other datasets, a 70/30 split is applied for training and validation, respectively; the RWKV model is initialized with weights from ~\cite{duan2024vision}.

\subsection{Comparison with State-of-the-Art Methods}
\label{subsec:comparison}

We compare our U-RWKV model against several state-of-the-art methods.
Table~\ref{Tab1:on5datasets} presents the Dice scores on five datasets, along with comparisons of the number of parameters (in M) and FLOPs (in G) for different models, which reflect computational complexity. Our U-RWKV model achieves competitive performance, attaining the highest average Dice score of \textbf{82.27}, surpassing most existing methods. Notably, the lightweight variant U-RWKV-s achieves a Dice score of 80.06 with only 0.46M parameters, highlighting its efficiency.

\begin{table*}[!t]
    \caption{Segmentation performances of competing methods in terms of Dice score. Reported Params, GFLOPs, average Dice (Avg), and Dice scores per dataset. Higher Dice values are better. Maximum values are highlighted in {\bf bold}.}
    \label{Tab1:on5datasets}
    \begin{center}
        \resizebox{\textwidth}{!}{%
            \begin{tabular}{l|cc|c|ccccc}
                \hline
                Methods & params & FLOPs & Avg & BUSI & Kvasir & Clinic & ISIC'17 & ISIC'18 \\
                \hline
                U-Net~\cite{ronneberger2015unet} & 34.53 & 65.52 & 81.48 & 79.58 & 87.65 & 90.97 & 89.87 & 86.99 \\
                TransUnet~\cite{chen2021transunet} & 93.23 & 32.23 & 81.23 & 79.61 & 87.13 & 90.84 & 90.10 & 86.58 \\
                CMU-Net~\cite{tang2023cmu} & 49.93 & 91.25 & 83.06 & 81.92 & \textbf{89.12} & \textbf{92.48} & 89.70 & 86.83 \\                
                SwinUnet~\cite{cao2022swin} & 27.15 & 5.91 & 76.05 & 76.46 & 80.67 & 84.15 & 87.71 & 86.57 \\
                \midrule
                UNeXt~\cite{valanarasu2022unext} & 1.47 & \textbf{0.57} & 80.18 & 80.47 & 85.11 & 88.76 & 89.60 & 86.80 \\
                Att-UNet~\cite{oktay2018attentionunet} & 4.91 & 9.45 & 81.48 & 79.61 & 87.13 & 91.77 & 89.57 & 86.86 \\
                MedT~\cite{valanarasu2021medical} & 1.37 & 2.41 & 81.81 & 81.86 & 88.85 & 90.38 & 86.72 & 87.21 \\
                ConvUNeXt~\cite{han2022convunext} & 3.51 & 7.25 & 81.11 & 80.37 & 86.67 & 90.99 & 89.35 & 85.89 \\
                CMUNeXt~\cite{tang2024cmunext} & 3.14 & 7.41 & 82.13 & 81.66 & 87.82 & 91.21 & 89.85 & 86.77 \\
                TinyUnet~\cite{chen2024tinyu} & \underline{0.48} & 1.67 & 80.45 & 77.42 & 87.32 & 90.37 & 89.03 & 86.87 \\
                U-RWKV-s & \textbf{0.46} & \underline{1.02} & 80.06 & 79.77 & 86.15 & 87.98 & 89.41 & 86.94 \\
                U-RWKV & 2.97 & 7.28 & \textbf{82.27} & \textbf{82.34} & 88.17 & 90.58 & \textbf{90.13} & \textbf{87.26} \\
                \hline
            \end{tabular}
        }
    \end{center}
\end{table*}

\begin{figure}[t!]
    \centering
    \includegraphics[width=0.94\linewidth]{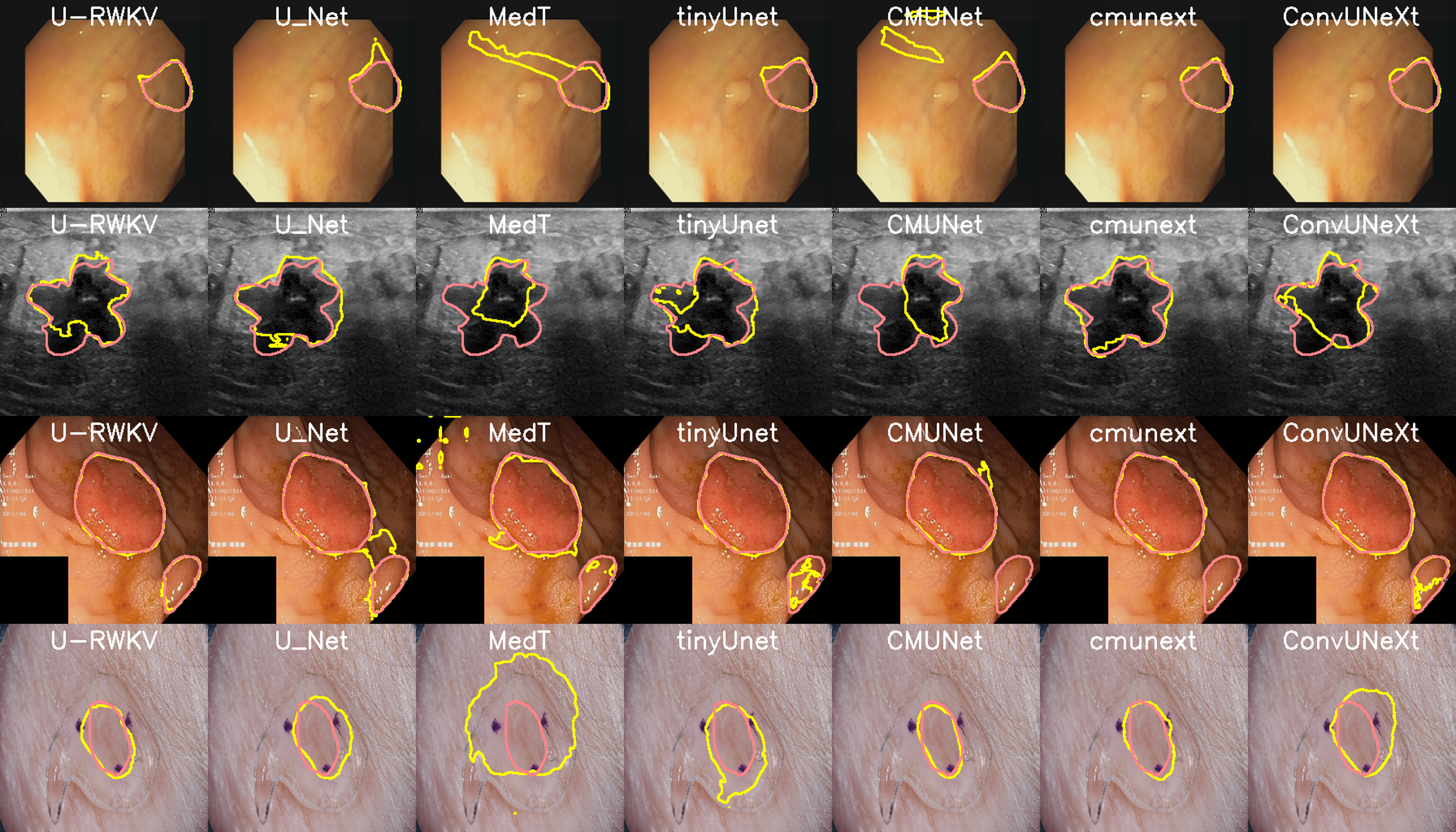}
    \caption{Visualization of segmentation results on four datasets: CVC-ClinicDB, BUSI, Kvasir-Seg, and ISIC 2017. The orange contour lines represent the ground-truth annotations of lesions, while the yellow contour lines indicate the segmentation results produced by our model.}
    \label{fig:visualization}
\end{figure}

We evaluate U-RWKV on the Synapse multi-organ segmentation dataset. Table~\ref{Tab2:synapse_} shows the Dice scores and Hausdorff Distance (HD95) for each organ, along with the average scores across all organs. Our U-RWKV model achieves competitive performance, with an average Dice score of \textbf{80.64} and HD95 of {26.61}. Notably, U-RWKV outperforms several state-of-the-art methods, including TransUNet and MedT. 

The synergy between RWKV’s long-range dependency modeling and SASE’s stage-adaptive feature refinement is key to this performance. SASE dynamically enhances lesion-specific features at different decoder stages—coarse-grained localization in early stages and fine-grained boundary precision in later ones. This is evident in Fig.~\ref{fig:visualization}, where U-RWKV reliably handles diverse challenges: heterogeneous textures in BUSI, mucosal folds in Kvasir, and low-contrast boundaries in ISIC. The ground truth (orange) and predictions (yellow) align closely, particularly for irregular structures, highlighting SASE’s role in preserving topological consistency.

\begin{table*}[!h]
    \centering
    \renewcommand{\arraystretch}{1.1} 
    {\fontsize{8}{9.5}\selectfont     
    \begin{minipage}{0.42\textwidth}
        \centering
        \caption{Comparison of different models on Synapse dataset.}
        \label{Tab2:synapse_}
        \begin{tabular}{l|cc}
            \toprule
            Methods & Dice$\uparrow$ & HD95$\downarrow$ \\
            \midrule
            U-Net              & 77.10 & 29.97 \\
            TransUNet          & 77.54 & 38.78 \\
            CMU-Net             & 76.22 & 29.65 \\
            \midrule
            UNeXt              & 72.52 & 39.61 \\
            AttUNet            & 76.10 & 36.77 \\
            MedT               & 70.09 & 33.53 \\
            ConvUNeXt          & \underline{78.55} & 26.89 \\
            CMUNeXt            & 77.95 & \textbf{24.43} \\
            TinyUnet           & 75.75 & 32.23 \\
            \midrule
            U-RWKV-s           & 71.12 & 45.36 \\
            U-RWKV             & \textbf{80.64} & \underline{26.61} \\ 
            \bottomrule
        \end{tabular}
    \end{minipage}
    \hfill
    \begin{minipage}{0.56\textwidth}
        \centering
        \caption{IoU results of ablation studies on BUSI, Kvasir and ISIC'17.}
        \label{tab3_:ablation_Dice}
        \begin{tabular}{l|c|c|c|c}
            \toprule
            Ablation & BUSI$\uparrow$  & Kvasir$\uparrow$ & ISIC$\uparrow$ & Avg.$\uparrow$ \\
            \midrule
            \textbf{L}eft$\rightarrow$\textbf{R}ight        & 69.61 & 77.91 & 81.65 & 76.75 \\
            \textbf{R}ight$\rightarrow$\textbf{L}eft        & 69.08 & 77.56 & 82.26 & 76.61 \\
            \textbf{T}op$\rightarrow$\textbf{B}ottom        & 69.04 & 78.87 & 82.23 & 77.02 \\
            \textbf{B}ottom$\rightarrow$\textbf{T}op        & 68.05 & 78.22 & 81.77 & 76.43 \\
            L.$\rightarrow$R.+R.$\rightarrow$L. & 69.73 & 77.22 & 81.90 & 76.64 \\
            T.$\rightarrow$B.+B.$\rightarrow$T.  & 69.73 & 78.35 & 81.98 & 76.98 \\
            L.$\rightarrow$R.+T.$\rightarrow$B. & 69.38 & 77.74 & 82.26 & 76.77 \\
            R.$\rightarrow$L.+B.$\rightarrow$T. & 68.86 & 78.75 & 81.92 & 76.86 \\
            w/o Dual RWKV                  & 70.30 & 77.31 & 82.42 & 76.68 \\
            w/o DARM                       & 66.73 & 76.55 & 81.62 & 74.97 \\
            w/o SASE                       & 68.57 & 75.78 & 81.53 & 75.29 \\
            U-RWKV                         & \textbf{71.01} & \textbf{79.58} & \textbf{82.27} & \textbf{77.62} \\
            \bottomrule
        \end{tabular}
    \end{minipage}
    }
\end{table*}

\subsection{Ablation Studies}
\label{AblationStudies}
We conduct comprehensive ablation studies, summarized in Table~\ref{tab3_:ablation_Dice}. The baseline results show that combining multi-directional scans (e.g., L$\rightarrow$R + R$\rightarrow$L, T$\rightarrow$B + B$\rightarrow$T, etc.) improves IoU across datasets. Removing components such as Dual RWKV or DARM causes notable performance drops (e.g., without DARM, the average IoU decreases from 77.62 to 74.97). Importantly, the full U-RWKV, which integrates SASE with DARM, achieves the highest average IoU of 77.62, demonstrating that SASE on its own is not merely a variant but works synergistically with DARM to enhance segmentation performance. 

We further validate these findings through Effective Receptive Field analysis. As shown in Fig.~\ref{fig:ERF_and_scatter}(b), U-RWKV achieves a 0.992 high-contribution area ratio at 0.99 threshold, significantly outperforming both the baseline U-Net (0.568) and our model's variants with ViT~\cite{dosovitskiy2020image} or VMUnet's~\cite{ruan2024vmunetvisionmambaunet}mamba bottlenecks (both 0.946). This 74.6\% improvement over U-Net and 4.9\% advantage over the backbone variants confirms our architecture's superior ability to focus activation energy on diagnostically relevant regions while maintaining global context awareness, consistent with the IoU improvements observed in the component ablation studies.

\section{Conclusion}
In summary, we propose U-RWKV, a framework that combines convolutional features with DARM's global dependency modeling, achieving a good balance between efficiency and accuracy. We acknowledge that inference speed is slightly slower than CNNs like UNeXt, and future work will focus on optimizing for high-resolution settings and extending to 3D segmentation.

\begin{credits}
\subsubsection{\ackname} Supported by National Natural Science Foundation of China under Grant 62271465, Suzhou Basic Research Program under Grant SYG202338.

\subsubsection{\discintname}
The authors have no competing interests to declare that are
relevant to the content of this article.
\end{credits}

\bibliographystyle{splncs04.bst}
\bibliography{reference}

\begin{thebibliography}{10}
\providecommand{\url}[1]{\texttt{#1}}
\providecommand{\urlprefix}{URL }
\providecommand{\doi}[1]{https://doi.org/#1}

\bibitem{al2020dataset}
Al-Dhabyani, W., Gomaa, M., Khaled, H., Fahmy, A.: Dataset of breast ultrasound images. Data in brief  \textbf{28},  104863 (2020)

\bibitem{bernal2015wm}
Bernal, J., S{\'a}nchez, F.J., Fern{\'a}ndez-Esparrach, G., Gil, D., Rodr{\'\i}guez, C., Vilari{\~n}o, F.: Wm-dova maps for accurate polyp highlighting in colonoscopy: Validation vs. saliency maps from physicians. Computerized medical imaging and graphics  \textbf{43},  99--111 (2015)

\bibitem{cao2022swin}
Cao, H., Wang, Y., Chen, J., Jiang, D., Zhang, X., Tian, Q., Wang, M.: Swin-unet: Unet-like pure transformer for medical image segmentation. In: European conference on computer vision. pp. 205--218. Springer (2022)

\bibitem{chen2021transunet}
Chen, J., Lu, Y., Yu, Q., Luo, X., Adeli, E., Wang, Y., Lu, L., Yuille, A.L., Zhou, Y.: Transunet: Transformers make strong encoders for medical image segmentation. arXiv preprint arXiv:2102.04306  (2021)

\bibitem{chen2024tinyu}
Chen, J., Chen, R., Wang, W., Cheng, J., Zhang, L., Chen, L.: Tinyu-net: Lighter yet better u-net with cascaded multi-receptive fields. In: MICCAI. pp. 626--635. Springer (2024)

\bibitem{codella2019skin}
Codella, N., Rotemberg, V., Tschandl, P., Celebi, M.E., Dusza, S., Gutman, D., Helba, B., Kalloo, A., Liopyris, K., Marchetti, M., et~al.: Skin lesion analysis toward melanoma detection 2018: A challenge hosted by the international skin imaging collaboration (isic). arXiv preprint arXiv:1902.03368  (2019)

\bibitem{ding2022scaling}
Ding, X., Zhang, X., Han, J., Ding, G.: Scaling up your kernels to 31x31: Revisiting large kernel design in cnns. In: CVPR. pp. 11963--11975 (2022)

\bibitem{dosovitskiy2020image}
Dosovitskiy, A., Beyer, L., Kolesnikov, A., Weissenborn, D., Zhai, X., Unterthiner, T., Dehghani, M., Minderer, M., Heigold, G., Gelly, S., et~al.: An image is worth 16x16 words: Transformers for image recognition at scale. arXiv preprint arXiv:2010.11929  (2020)

\bibitem{duan2024vision}
Duan, Y., Wang, W., Chen, Z., Zhu, X., Lu, L., Lu, T., Qiao, Y., Li, H., Dai, J., Wang, W.: Vision-rwkv: Efficient and scalable visual perception with rwkv-like architectures. arXiv preprint arXiv:2403.02308  (2024)

\bibitem{guo2020polyp}
Guo, Y., Bernal, J., J~Matuszewski, B.: {Polyp Segmentation with Fully Convolutional Deep Neural Networks—Extended Evaluation Study}. Journal of Imaging  \textbf{6}(7), ~69 (2020)

\bibitem{han2022convunext}
Han, Z., Jian, M., Wang, G.G.: Convunext: An efficient convolution neural network for medical image segmentation. Knowledge-based systems  \textbf{253},  109512 (2022)

\bibitem{huang2020unet3plus}
Huang, H., Lin, L., Tong, R., Hu, H., Zhang, Q., Iwamoto, Y., Han, X., Chen, Y.W., Wu, J.: Unet 3+: A full-scale connected unet for medical image segmentation. In: ICASSP. pp. 1055--1059. IEEE (2020)

\bibitem{hz2-huang2024pele}
Huang, Z., Li, H., Shao, S., Zhu, H., Hu, H., Cheng, Z., Wang, J., Kevin~Zhou, S.: Pele scores: pelvic x-ray landmark detection with pelvis extraction and enhancement. IJCAS  \textbf{19}(5),  939--950 (2024)

\bibitem{Jhadoubleunet2020}
Jha, D., Riegler, M., Johansen, D., Halvorsen, P., Johansen, H.: {DoubleU-Net: A Deep Convolutional Neural Network for Medical Image Segmentation}. In: CBMS (2020)

\bibitem{jha2019resunet++}
Jha, D., Smedsrud, P.H., Riegler, M.A., Johansen, D., De~Lange, T., Halvorsen, P., Johansen, H.D.: {ResUNet++: An Advanced Architecture for Medical Image Segmentation}. In: ISM. pp. 225--230 (2019)

\bibitem{luo2016understanding}
Luo, W., Li, Y., Urtasun, R., Zemel, R.: Understanding the effective receptive field in deep convolutional neural networks. NeurIPS  \textbf{29} (2016)

\bibitem{oktay2018attentionunet}
Oktay, O.: Attention u-net: Learning where to look for the pancreas. arXiv preprint arXiv:1804.03999  (2018)

\bibitem{peng2023rwkvcem}
Peng, B., Alcaide, E., Anthony, Q.G., Albalak, A., Arcadinho, S., Biderman, S., Cao, H., Cheng, X., Chung, M.N., Derczynski, L., et~al.: Rwkv: Reinventing rnns for the transformer era. In: EMNLP (2023)

\bibitem{richardson2022nature}
Richardson, S., Lawrence, K., Schoenthaler, A.M., Mann, D.: A framework for digital health equity. NPJ digital medicine  \textbf{5}(1), ~119 (2022)

\bibitem{ronneberger2015unet}
Ronneberger, O., Fischer, P., Brox, T.: U-net: Convolutional networks for biomedical image segmentation. In: Medical image computing and computer-assisted intervention--MICCAI 2015: 18th international conference, Munich, Germany, October 5-9, 2015, proceedings, part III 18. pp. 234--241. Springer (2015)

\bibitem{ruan2024vmunetvisionmambaunet}
Ruan, J., Li, J., Xiang, S.: Vm-unet: Vision mamba unet for medical image segmentation (2024), \url{https://arxiv.org/abs/2402.02491}

\bibitem{tang2024cmunext}
Tang, F., Ding, J., Quan, Q., Wang, L., Ning, C., Zhou, S.K.: Cmunext: An efficient medical image segmentation network based on large kernel and skip fusion. In: 2024 IEEE International Symposium on Biomedical Imaging (ISBI). pp.~1--5. IEEE (2024)

\bibitem{tang2023mobileutr}
Tang, F., Nian, B., Ding, J., Quan, Q., Yang, J., Liu, W., Zhou, S.K.: Mobileutr: Revisiting the relationship between light-weight cnn and transformer for efficient medical image segmentation. arXiv preprint arXiv:2312.01740  (2023)

\bibitem{tang2025mambamim}
Tang, F., Nian, B., Li, Y., Jiang, Z., Yang, J., Liu, W., Zhou, S.K.: Mambamim: Pre-training mamba with state space token interpolation and its application to medical image segmentation. Medical Image Analysis p. 103606 (2025)

\bibitem{tang2023cmu}
Tang, F., Wang, L., Ning, C., Xian, M., Ding, J.: Cmu-net: a strong convmixer-based medical ultrasound image segmentation network. In: ISBI. pp.~1--5. IEEE (2023)

\bibitem{tang2024hyspark}
Tang, F., Xu, R., Yao, Q., Fu, X., Quan, Q., Zhu, H., Liu, Z., Zhou, S.K.: Hyspark: Hybrid sparse masking for large scale medical image pre-training. In: MICCAI. pp. 330--340. Springer (2024)

\bibitem{tang2025hi}
Tang, F., Yao, Q., Ma, W., Wu, C., Jiang, Z., Zhou, S.K.: Hi-end-mae: Hierarchical encoder-driven masked autoencoders are stronger vision learners for medical image segmentation. arXiv preprint arXiv:2502.08347  (2025)

\bibitem{valanarasu2021medical}
Valanarasu, J.M.J., Oza, P., Hacihaliloglu, I., Patel, V.M.: Medical transformer: Gated axial-attention for medical image segmentation. In: MICCAI. pp. 36--46 (2021)

\bibitem{valanarasu2022unext}
Valanarasu, J.M.J., Patel, V.M.: Unext: Mlp-based rapid medical image segmentation network. In: MICCAI. pp. 23--33. Springer (2022)

\bibitem{hz3-zhou2024hybrid}
Zhou, X., Huang, Z., Zhu, H., Yao, Q., Zhou, S.K.: Hybrid attention network: An efficient approach for anatomy-free landmark detection. arXiv preprint arXiv:2412.06499  (2024)

\bibitem{zhou2018unet++}
Zhou, Z., Rahman~Siddiquee, M.M., Tajbakhsh, N., Liang, J.: Unet++: A nested u-net architecture for medical image segmentation. In: Deep Learning in Medical Image Analysis and Multimodal Learning for Clinical Decision Support: 4th International Workshop, DLMIA 2018, and 8th International Workshop, ML-CDS 2018, Held in Conjunction with MICCAI 2018, Granada, Spain, September 20, 2018, Proceedings 4. pp. 3--11. Springer (2018)

\end{thebibliography}

\end{document}